\documentclass[12pt,a4paper]{article}
\usepackage{cite}
\begin{document}
\title{\bf Ensemble teleportation}
\author{{\bf Thomas Kr\"uger}\thanks{E-mail: t.krueger@phys.uni-paderborn.de}\\Theoretical Physics, Faculty of Science, University of Paderborn\\Warburger Str. 100, 33098 Paderborn, Germany}
\date{}
\maketitle
\begin{abstract}
The possibility of teleportation is by sure the most interesting consequence of quantum non-separability. So far, however, teleportation schemes have been formulated by use of state vectors and considering individual entities only. In the present article the feasibility of teleportation is examined on the basis of the rigorous ensemble interpretation of quantum mechanics (not to be confused with a mere treatment of noisy EPR pairs) leading to results which are unexpected from the usual point of view.

\end{abstract}

\newpage
\section{Introduction}
The possibility of teleportation is by sure the most interesting consequence of quantum non-separability. Bennett et al. were the first who have realized that an "unknown quantum state $|\phi\rangle$ can be disassembled into, then later reconstructed from, purely classical information and purely nonclassical Einstein-Podolsky-Rosen (EPR) correlations. To do so the sender, 'Alice,' and the receiver, 'Bob,' must prearrange the sharing of an EPR-correlated pair of particles. Alice makes a joint measurement on her EPR particle and the unknown quantum system, and sends Bob the classical result ... Knowing this, Bob can convert the state of his EPR particle into an exact replica of the unknown state ... which Alice destroyed." \cite{a4}. Meanwhile several experiments have been performed successfully by different research groups, thereby demonstrating the feasibility of teleportation \cite{a5,a6,a7,a8,a9,a10}. With the exception of \cite{a10} all experiments employ photons both for the generation of the EPR pair and to materialize the unknown state. There is a couple of proposals to realize teleportation using massive entities, i. e., atoms in high-\emph{Q} cavities \cite{a11,a12,a13,a14}, solid-state systems \cite{a15}, and clouds of atoms \cite{a16,a17}. A crucial point of every experimental implementation of a theoretical teleportation scheme is the joint measurement of Alice's half of the EPR pair and the unknown entity the state of which shall be teleported \cite{a18}. At present new ideas are discussed to overcome these difficulties \cite{a19,a20,a21}, and numerous researchers investigate the teleportation fidelity if non-ideal EPR pairs are used (see, e. g., \cite{a22,a23,a24,a25}).

So far teleportation schemes have been formulated by use of state vectors and considering individual entities only. In previous articles the present author has shown that an \emph{ensemble} interpretation of quantum mechanics (QM) is useful to unveil precisely the actual core of EPR's discovery \cite{a1,a2,a3} which is the non-separability of the quantum world. So it is a challenging task to analyze teleportation in the frame of said ensemble interpretation as well, especially since some of the proposals mentioned above seem to make explicit use of ensembles. [Note, however, that an ensemble in the strict sense of the word is \emph{not} a many-particle system. There the constituents always interact to a certain extent, but an ensemble is, at least for all practical purposes, interaction-free, i. e., it can be thought of as a sufficiently diluted many-particle system.] Therefore statistical operators are used instead of state vectors, because they reflect directly the ensemble way of viewing QM. Recall that the founding fathers of QM were convinced that the state vectors refer to individual entities only.

In the following we will \emph{not} make use of the so-called simple ensemble interpretation which rests on the idea that each individual member of the ensemble always \emph{has} (in the sense of EPR's principle of reality) precise values (properties) for all its property types. Gillespie has proven that, in the case of a certain quantum system, this interpretation is not possible \cite{a31}. This result, however, is not of relevance for the approach employed in the present article, because here no statements regarding properties to be ascribed to individual entities are made. Only the whole, i. e., the ensemble, will be considered as an element of QM.

It should have become clear that ensembles in the sense described above have nothing in common with the so-called noisy EPR pairs often discussed in the context of non-ideal teleportation.
\section{The teleportation process}
\subsection{Preliminary remarks}
Basis of all following considerations is an \emph{ensemble} interpretation of QM, i. e., it is presupposed that QM makes
\begin{itemize}
\item statistical statements on
\item the results of measurements on
\item ensembles.
\end{itemize}
This approach offers a couple of advantages in handling of non-separability and quantum holism which are explained in detail in \cite{a1,a2,a3}.

A given ensemble of micro-entities is represented by a self-adjoint operator $\rho$, the so-called statistical operator, with $\mathrm{Tr}(\rho) = 1$ and positive spectrum (including 0). Now let this $\rho$ be defined on a $2^3$ dimensional Hilbert space $\mathcal{H}_{total} = \mathcal{H}_A \otimes \mathcal{H}_B \otimes \mathcal{H}_C$ where the set $\{|\alpha_1 \rangle,|\alpha_2 \rangle\}$ forms an orthonormal basis of the subspace $\mathcal{H}_A$, and let respective basis sets be given analogously for the other subspaces. In the four dimensional subspace $\mathcal{H}_A \otimes \mathcal{H}_B$ we generate a Bell-type basis according to
\begin{eqnarray}
|\Psi_e^\pm \rangle : = \frac{1}{\sqrt{2}}\: (|\alpha_1\rangle |\beta_1\rangle \pm |\alpha_2\rangle |\beta_2\rangle)\\
|\Psi_o^\pm \rangle : = \frac{1}{\sqrt{2}}\: (|\alpha_1\rangle |\beta_2\rangle \pm |\alpha_2\rangle |\beta_1\rangle).
\end{eqnarray}
With the aid of these Bell-basis four statistical operators can be defined,
\begin{eqnarray}
\rho_{1,2}& :=& |\Psi_e^\pm\rangle \: \langle\Psi_e^\pm| \nonumber \\
&=& \frac{1}{2} \: (\hat{A}_{11} \otimes \hat{B}_{11} \pm \hat{A}_{12} \otimes \hat{B}_{12} \pm \hat{A}_{21} \otimes \hat{B}_{21} + \hat{A}_{22} \otimes \hat{B}_{22}) \\
\rho_{3,4}& :=& |\Psi_o^\pm\rangle \: \langle\Psi_o^\pm| \nonumber \\
&=& \frac{1}{2} \: (\hat{A}_{11} \otimes \hat{B}_{22} \pm \hat{A}_{12} \otimes \hat{B}_{21} \pm \hat{A}_{21} \otimes \hat{B
}_{12} + \hat{A}_{22} \otimes \hat{B}_{11}),
\end{eqnarray}
where $\hat{A}_{ij} = |\alpha_i\rangle \: \langle\alpha_j|$ and $\hat{B}_{ij} = |\beta_i\rangle \: \langle\beta_j|$. For these four operators the following statements are valid:
\begin{itemize}
\item $\rho_i^2 = \rho_i$
\item $\rho_i \rho_j = \hat{0}\: \forall\: i \ne j$
\item They are \emph{non}-separable \cite{a2}.
\end{itemize}
\subsection{A teleportation scheme for ensembles}
\subsubsection{Formalism}
Suppose we are in possession of a generator producing an ensemble \{(AB)$_i$\} of micro-entities (AB)$_i$ each of them dissociating according to (AB)$_i \longrightarrow$ A$_i$ + B$_i$. The sub-ensemble of all A$_i$ is sent either one by one or as a whole to an observer named Alice whereas the sub-ensemble of all B$_i$ is sent correspondingly to a second observer named Bob. [The problem of storing ensembles is discussed in section 3.] The ensemble \{\{A$_i$\},\{B$_i$\}\} consisting of the two sub-ensembles is represented, say, by the statistical operator $\rho_4$ (see eq. 4) which means that we consider the ensemble in a \emph{pure} state.

Now Alice obtains a further entity-ensemble, \{C$_i$\}, which is assumed to be in a pure state as well. Then the new total ensemble \{\{C$_i$\},\{A$_i$\},\{B$_i$\}\} is represented by
\begin{equation}
\rho_{total} = \rho_C \otimes \rho_4
\end{equation}
where $\rho_C$ is given by
\begin{equation}
\rho_C = \sum_{k,l=1}^2 c_{kl} \hat{C}_{kl}.
\end{equation}
Tr$(\rho_C) = 1$. The actual values of the coefficients $c_{kl}$ must not be known to Alice.
\begin{eqnarray}
\Rightarrow \rho_{total}& =& \frac{1}{2} \: (c_{11} \hat{C}_{11} + c_{12} \hat{C}_{12} + c_{21} \hat{C}_{21} + c_{22} \hat{C}_{22})\nonumber\\& & \otimes \: (\hat{A}_{11} \otimes \hat{B}_{22} - \hat{A}_{12} \otimes \hat{B}_{21} - \hat{A}_{21} \otimes \hat{B}_{12} + \hat{A}_{22} \otimes \hat{B}_{11})
\end{eqnarray}
We define four new statistical operators, $\rho'_1$, $\rho'_2$, $\rho'_3$, and $\rho'_4$, so that they are analogous to the already introduced operators $\rho_1$, $\rho_2$, $\rho_3$, and $\rho_4$, respectively. The primed operators, however, shall act on $\mathcal{H}_C \otimes \mathcal{H}_A$ instead of $\mathcal{H}_A \otimes \mathcal{H}_B$. With the aid of these new operators, and after some lengthy but straightforward manipulations, $\rho_{total}$ can be brought into the form
\begin{eqnarray}
2 \rho_{total} &=& \rho'_1 \otimes (c_{11} \hat{B}_{22} - c_{12} \hat{B}_{21} - c_{21} \hat{B}_{12} + c_{22} \hat{B}_{11})\nonumber\\& &+\:\rho'_2 \otimes (c_{11} \hat{B}_{22} + c_{12} \hat{B}_{21} + c_{21} \hat{B}_{12} + c_{22} \hat{B}_{11})\nonumber\\& &+\:\rho'_3 \otimes (c_{11} \hat{B}_{11} - c_{12} \hat{B}_{12} - c_{21} \hat{B}_{21} + c_{22} \hat{B}_{22})\nonumber\\& &+\:\rho'_4 \otimes (c_{11} \hat{B}_{11} + c_{12} \hat{B}_{12} + c_{21} \hat{B}_{21} + c_{22} \hat{B}_{22})\nonumber\\& &+\:[\ldots]
\end{eqnarray}
where
\begin{equation}
[\ldots] = \sum_{i,j=1}^2 \hat{F}_{ij}
\end{equation}
and
\begin{eqnarray}
\hat{F}_{11} &=& -\:(c_{22} \hat{A}_{11} \otimes \hat{B}_{11} + c_{11} \hat{A}_{12} \otimes \hat{B}_{21}\nonumber\\ & & +\: c_{11} \hat{A}_{21} \otimes \hat{B}_{12} + c_{22} \hat{A}_{22} \otimes \hat{B}_{22}) \otimes \hat{C}_{11} \\ \hat{F}_{12} &=& (c_{12} \hat{A}_{11} \otimes \hat{B}_{22} + c_{21} \hat{A}_{12} \otimes \hat{B}_{12}\nonumber\\ & & +\: c_{21} \hat{A}_{21} \otimes \hat{B}_{21} + c_{12} \hat{A}_{22} \otimes \hat{B}_{11}) \otimes \hat{C}_{12}
\end{eqnarray}
\begin{eqnarray}
\hat{F}_{21} &=& (c_{21} \hat{A}_{11} \otimes \hat{B}_{22} + c_{12} \hat{A}_{12} \otimes \hat{B}_{12}\nonumber\\ & & +\: c_{12} \hat{A}_{21} \otimes \hat{B}_{21} + c_{21} \hat{A}_{22} \otimes \hat{B}_{11}) \otimes \hat{C}_{21} \\ \hat{F}_{22} &=& -\:(c_{11} \hat{A}_{11} \otimes \hat{B}_{11} + c_{22} \hat{A}_{12} \otimes \hat{B}_{21}\nonumber\\ & & +\: c_{22} \hat{A}_{21} \otimes \hat{B}_{12} + c_{11} \hat{A}_{22} \otimes \hat{B}_{22}) \otimes \hat{C}_{22}.
\end{eqnarray} 
By use of
\begin{equation}
\rho_B = c_{11} \hat{B}_{11} + c_{12} \hat{B}_{12} + c_{21} \hat{B}_{21} + c_{22} \hat{B}_{22}
\end{equation}
and employing the Pauli matrices $\sigma_1$ and $\sigma_3$, both related to the subspace $\mathcal{H}_B$, eq. 8 can be written as
\begin{eqnarray}
2 \rho_{total} &=& \rho'_1 \otimes (\sigma_3 \sigma_1 \rho_B \sigma_1 \sigma_3) + \rho'_2 \otimes (\sigma_1 \rho_B \sigma_1) + \rho'_3 \otimes (\sigma_3 \rho_B \sigma_3)\nonumber\\ & & + \:\rho'_4 \otimes \rho_B +\: [\ldots].
\end{eqnarray}
Now, in the usual teleportation scheme formulated on the basis of wavefunctions of individual objects, Alice projects by by more or less tricky operations her combined system C+A onto one of the four states $|\Phi_e^{\pm}\rangle = \frac{1}{\sqrt2} \left(|\gamma_1\rangle |\alpha_1\rangle \pm |\gamma_2\rangle |\alpha_2\rangle \right)$ and $|\Phi_o^{\pm}\rangle = \frac{1}{\sqrt2} \left(|\gamma_1\rangle |\alpha_2\rangle \pm |\gamma_2\rangle |\alpha_1\rangle\right)$, respectively. This operation, however, changes the state of the total system C+A+B too. We will see that it is not necessary for Alice to take note of the outcome of the projection. It is sufficient to transmit the information about what she has \emph{done}. Employing a classical transmission channel Bob learns what to do in order to transform the state of his B into the original state of C. In this way a state can be teleported.
\subsubsection{Interpretation}
In the case of ensembles things might be quite different. Assume for the moment that Alice's operation on her sub-ensemble \{\{C$_i$\},\{A$_i$\}\} consists of the projection onto $\rho'_1$, i. e., the preparation process is described by
\begin{equation}
\mathrm{Tr_{C,A}}\left((\hat{P} \otimes \hat{1}_B) \rho_{total} \right)
\end{equation}
where $\hat{P} = \rho'_1$ (see Appendix). Due to the mutual orthogonality of the primed operators we obtain
\begin{eqnarray}
\mathrm{Tr_{C,A}}\left((\rho'_1 \otimes \hat{1}_B) \rho_{total} \right) &=& \frac{1}{4} \: (-c_{22} \hat{B}_{11} + c_{21} \hat{B}_{12} + c_{12} \hat{B}_{21} - c_{11} \hat{B}_{22})\nonumber\\ & & + \:\frac{1}{2} \: \sigma_3 \sigma_1 \rho_B \sigma_1 \sigma_3.
\end{eqnarray}
Insertion of $\rho_B$ from eq. 14 yields
\begin{equation}
\sigma_3 \sigma_1 \rho_B \sigma_1 \sigma_3 = c_{22} \hat{B}_{11} - c_{21} \hat{B}_{12} - c_{12} \hat{B}_{21} + c_{11} \hat{B}_{22}.
\end{equation}
\begin{equation}
\Rightarrow \mathrm{Tr_{C,A}}\left((\rho'_1 \otimes \hat{1}_B) \rho_{total} \right) = \frac{1}{4} \: \sigma_3 \sigma_1 \rho_B \sigma_1 \sigma_3
\end{equation}
It is seen immediately that this operator has a trace of 1/4, i. e., it must be re-normalized by division by its norm.
\begin{equation}
\Rightarrow \mathrm{\tilde{Tr}_{C,A}}\left((\rho'_1 \otimes \hat{1}_B) \rho_{total} \right) = \sigma_3 \sigma_1 \rho_B \sigma_1 \sigma_3
\end{equation}
This is the result of Alice's operations, and this is what Bob has at hand. Now Alice must tell Bob what she has done, i. e., if, before the very beginning of the whole experiment, Alice and Bob have agreed upon both the allowed preparations on Alice's side and a corresponding number code, then Alice has to send two bits of classical information to Bob, and he will be able to apply some operations so that his sub-ensemble \{B$_i$\} is represented by $\rho_B$ which is the analog of the unknown $\rho_C$ in Bob's Hilbert space. In this way teleportation of a statistical operator is achieved, but recall that Alice must inform Bob about what she has \emph{done}, not about the \emph{result} of her action, i. e., Alice does not read any pointer position. She makes a pure preparation without any gain or loss of information regarding the ensemble! The only thing one has to require is that the operator representing what Alice does must be physically realizable.

One could object that, in the case of ensembles, teleportation looses its unique quantum features because Alice simply could sacrifice some of the identical constituents of the C-ensemble to determine $\rho_C$ (eq. 6) and then transmit the information to Bob via a classical channel. She could, at least in principle, perform a measurement even of the \emph{complete} C-ensemble, therefore obtaining \emph{precise} values for all coefficients $c_{kl}$ and, by transmitting a nevertheless finite number of bits, enable Bob to transform his B-sub-ensemble accordingly. This is trivial indeed. The point, however, is that Alice does not need to know the $c_{kl}$. She manipulates the A$\otimes$C-ensemble and, without knowing anything about C, still enables Bob to recreate it. And this is by sure non-trivial.
 
So far ensemble teleportation and the teleportation of individual micro-entities seem to be equivalent, and in fact, after application of the unitary transformations $\sigma_3 \ldots \sigma_3$ and $\sigma_1 \ldots \sigma_1$, Bob ends up with the desired result.
\subsection{Ensemble teleportation by use of one bit of information only}
In the most general case Alice's preparation consists in the application of the operator
\begin{equation}
\hat{P} = \sum_{k,l,m,n} u_{klmn} |\gamma_k\rangle \langle\gamma_l| \otimes |\alpha_m\rangle \langle\alpha_n| \equiv \sum_{k,l,m,n} u_{klmn} \:\hat{C}_{kl} \otimes \hat{A}_{mn}
\end{equation}
where $\sum_{k,m} u_{kkmm} = 1$ and $u_{kkmm} \ge 0 \:\forall\: k,m$. With $\rho_{total}$ from eq. 7 and using the fact that, e. g., $\hat{C}_{kl} \hat{C}_{pq} = \hat{C}_{kq}$ if $l=p$ and $\hat{0}$ otherwise we then obtain:
\begin{eqnarray}
(\hat{P} \otimes \hat{1}_B) \rho_{total} &=& \frac{1}{2} \sum_{k,m,p,q} c_{pq} \hat{C}_{kq} \otimes \left(\hat{A}_{m1} \otimes (u_{kpm1} \hat{B}_{22} - u_{kpm2} \hat{B}_{12})\right.\nonumber\\ & & + \: \hat{A}_{m2} \otimes \left.(- u_{kpm1} \hat{B}_{21} + u_{kpm2} \hat{B}_{11}) \right)
\end{eqnarray}
Again we form the trace with regard to both $\mathcal{H}_C$ and $\mathcal{H}_A$.
\begin{equation}
\Rightarrow \mathrm{Tr_{C,A}} \left((\hat{P} \otimes \hat{1}_B) \rho_{total} \right) = \frac{1}{2} \sum_{i,p} c_{pi} (u_{ip11} \hat{B}_{22} - u_{ip12} \hat{B}_{12} - u_{ip21} \hat{B}_{21} + u_{ip22} \hat{B}_{11})
\end{equation}
This equation defines an operator, called $\rho_{Bob}$, which, after re-normalization, is the result of Alice's preparation on her ensemble \{\{C$_i$\},\{A$_i$\}\}.
\begin{equation}
\tilde{\rho}_{Bob} = \frac{\rho_{Bob}}{\mathrm{Tr}(\rho_{Bob})}
\end{equation}
$\rho_{Bob}$ represents the essence of the actual state of Bob's sub-ensemble. In the basis of the operators $\hat{B}_{ij}$ it can be written vectorially as follows:
\begin{equation}
\rho_{Bob} = \frac{1}{2} \left( \begin{array}{r} c_{11} u_{1122} + c_{12} u_{2122} + c_{21} u_{1222} + c_{22} u_{2222} \\ -\: c_{11} u_{1112} - c_{12} u_{2112} - c_{21} u_{1212} - c_{22} u_{2212} \\ -\:c_{11} u_{1121} - c_{12} u_{2121} - c_{21} u_{1221} - c_{22} u_{2221} \\ c_{11} u_{1111} + c_{12} u_{2111} + c_{21} u_{1211} + c_{22} u_{2211} \end{array} \right)
\end{equation}
This vector results from the original vector $\vec{c} = (c_{11},c_{12},c_{21},c_{22})$ (see eq. 6) by the transformation
\begin{equation}
\mathbf{T} = \left( \begin{array}{*{4}{r}} u_{1122} & u_{2122} & u_{1222} & u_{2222} \\ -u_{1112} & -u_{2112} & -u_{1212} & -u_{2212} \\ -u_{1121} & -u_{2121} & -u_{1221} & -u_{2221} \\ u_{1111} & u_{2111} & u_{1211} & u_{2211} \end{array} \right)
,
\end{equation}
i. e.,
\begin{equation}
\tilde{\rho}_{Bob} = \frac{\frac{1}{2} \mathbf{T} \vec{c}}{\| \frac{1}{2} \mathbf{T} \vec{c} \, \|}.
\end{equation}
So Bob can impress the original state $\vec{c}$ on his sub-ensemble by applying the inverse transformation. Here it is presupposed that the inverse actually exists which is always the case if Alice projects on one of the $\rho'_i$.

But what would happen if Bob would completely renounce any manipulation of his $\tilde{\rho}_{Bob}$, i. e., if he would \emph{not} apply the inverse transformation? We start the investigation of this case by rewriting Bob's statistical operator in the form
\begin{equation}
\tilde{\rho}_{Bob} = \vec{c} + \left( \frac{ \mathbf{T}}{\| \mathbf{T} \vec{c} \, \|} - \mathbf{1}\right) \vec{c}
\end{equation}
where the second term represents the contamination of $\vec{c}$ due to the fact that Bob has not done anything at all. The success of a teleportation, the so-called fidelity $f$, is given by
\begin{equation}
f = \mathrm{Tr}(\rho_C|_{(ensemble B)} \times \tilde{\rho}_{Bob}).
\end{equation}
In terms of our matrix representation and by use of (28) this definition reduces to the sum of two scalar products:
\begin{equation}
f = \underbrace{\vec{c} \: \vec{c}}_{= 1} \; + \; \vec{c} \: \left( \frac{\mathbf{T}}{\| \mathbf{T} \vec{c} \, \|} - \mathbf{1} \right) \vec{c}.
\end{equation}
Now assume that Alice had performed a projection onto $\rho'_1$.
\begin{equation}
\Rightarrow \mathbf{T} = \left( \begin{array}{*{4}{r}} 0 & 0 & 0 & \frac{1}{2} \\ 0 & 0 & -\frac{1}{2} & 0 \\ 0 & -\frac{1}{2} & 0 & 0 \\ \frac{1}{2} & 0 & 0 & 0 \end{array} \right)
\end{equation}
\begin{eqnarray}
\Rightarrow \| \mathbf{T} \vec{c} \, \| &=& \frac{1}{2} \\ \Rightarrow f &=& 1 + \vec{c} \, (2 \mathbf{T} - \mathbf{1}) \, \vec{c} \\ &=& 2 c_{11} c_{22} - 2 c_{12} c_{21}
\end{eqnarray}
Since
\begin{itemize}
\item both $c_{11}$ and $c_{22} \ge 0$,
\item $c_{11} + c_{22} = 1$, and
\item $c_{12}c_{21} \le c_{11}c_{22} \le \frac{1}{4}$,
\end{itemize}
we can easily calculate the fidelity for some choice of these coefficients. The fidelity attains its maximum if both $c_{11}$ and $c_{22}$ are equal to $\frac{1}{2}$ and $|c_{12}|^2 = 0$. $\Rightarrow f_{max} = \frac{1}{2}$ which by sure is a poor result, i. e., laziness does not pay off.

Normally, however, $\rho_C$ is unknown and Bob must find ways to manipulate his sub-ensemble. Note that the average fidelity over all possible inputs has been determined, based on the usual teleportation scheme, to be $\frac{2}{3}$ \cite{a32}.
Alice, however, is in the position of being able to save some of Bob's efforts. Assume that Alice has already made up her mind about the details of her operation well before the beginning of the experiment, and assume further that she has told Bob what she wants to do. In this case Alice needs to send a single bit of classical information only to inform Bob that her part of the experiment has taken place. Strictly speaking, this bit can be reduced even to a single "ping" ($\equiv$ yes, it's done). But this idea can yet be carried on. If Alice and Bob, at the very beginning of the whole story, already achieve agreement on the \emph{time} when the preparation shall be performed, then Bob has to wait at that time only and he will then be able to reproduce $\vec{c}$ without obtaining any information from Alice at all! Note, however, that this idea applies to the usual teleportation scheme as well.

\subsection{Ensemble teleportation without any action on Bob's side}
Things become really fascinating if we choose
\begin{equation}
\mathbf{T} = \mathbf{1},
\end{equation}
because this immediately yields $\tilde{\rho}_{Bob} = \vec{c}$ (in the basis of the $\hat{B}_{ij}$), i. e., in this case Bob would \emph{have} the teleported state of $\{\mathrm{C}_i\}$ after Alice's preparation \emph{without doing anything at all}! This situation raises the following question:
Can the condition (35) imposed on $\mathbf{T}$ be realized by an operator $\hat{P}$? Then Alice would be able to prepare the total ensemble in a way that the \emph{unknown} state $\vec{c}$ is teleported \emph{automatically}.

From (35), (26), and (21) it is easy to see that the necessary operator for automatic teleportation is given by
\begin{equation}
\hat{P}_{aut} = \hat{C}_{11} \otimes \hat{A}_{22} - \hat{C}_{21} \otimes \hat{A}_{12} - \hat{C}_{12} \otimes \hat{A}_{21} + \hat{C}_{22} \otimes \hat{A}_{11}.
\end{equation}
The comparison of this equation and (4) shows that $\hat{P}_{aut} = 2 \: \rho'_4$, i. e., also $\hat{P}_{aut}$ is self-adjoint and its eigenvalues are $\lambda_{1,2} = + 1$ and $\lambda_{3,4} = - 1$. Proceeding in the same way as in the previous subsection we then obtain $\tilde{\rho}_{Bob} = \vec{c}$, i. e., complete ensemble teleportation is possible even if the second observer is a lazybone.

Let me repeat this strange result in plane words: Let Alice's measurement or operation be described by the operator $\hat{P}_{aut}$. After her manipulation of the C-A-ensemble Bob's B-sub-ensemble is in the state (28), i. e., in this case Bob's B-sub-ensemble would be in the desired state just after Alice's measurement - \emph{without} any action on Bob's side (as, e. g., selecting constituents of his sub-ensemble) and \emph{without} any information to be transmitted. At most Alice could send one single 'ping' if she has done her work so that Bob will know that his sub-ensemble is now in the state described by $\vec{c}$. But that is all. The only open question is how the operation given by $\hat{P}_{aut}$ can be realized in the lab. This question will be addressed in the following section.

This kind of teleportation of course has nothing to do with any mysterious effect as, e. g., superluminal communication. No miracle happens. The fundament of any EPR correlation is the fact that none of the two sub-ensembles has an independent existence of its own. Each of them remembers the common origin and is, therefore, in the possession of the \emph{full} information. This point is explained in detail in \cite{a1,a3}.

It is easy to see that, in contrast to the operators usually applied by Alice, 
\begin{equation}
\hat{P}_{aut}^2 = 2 \hat{P}_{aut},
\end{equation}
i. e., $\hat{P}_{aut}$ is no projection operator, and, on the other hand, it is also not an element of a positive operator-valued measure (POVM) because $\| \hat{P}_{aut} \| = 2$, but it nevertheless may represent a feasible action on an ensemble because it is self-adjoint as, e. g., the standard Hamiltonian operators as well.

In the case of automatic teleportation ($\mathbf{T} = \mathbf{1}$) (30) shows that the fidelity is equal to 1 irrespective of the incoming state $\vec{c}$. So an arbitrary incoming state can not only be transported without any action on Bob's side but also with maximum fidelity!

It should be mentioned that this proposal has nothing in common with what happens in the original experiment of the Zeilinger group \cite{a5}. In said experiment a projection onto one of the four Bell states has been realized insofar as a corresponding sub-ensemble was selected out of the original total ensemble. For this sub-ensemble, representing one quarter of the original constituents only, automatic teleportation was achieved. This approach, however, is totally different from what is proposed here. \emph{No} out-selection has to be performed. The \emph{whole} ensemble is present in the teleportation scheme. Furthermore, Alice's operation is $\hat{P}_{aut}$ and \emph{not} the projection onto $|\Psi^-_{12}\rangle$.
\section{Discussion}
\begin{itemize}
\item How do Alice and Bob perform their measurements/preparations? With individual entities this question is trivial, but how can \emph{ensemble} measurements be realized? If Alice obtains the constituents of her ensemble \{\{C$_i$\},\{A$_i$\}\} one by one, i. e., one pair C$_i$+A$_i$ at a time, then her activities do not differ from the usual case with the only exception that she has to process the whole ensemble until $\rho_{Bob}$ is available. But also this is not a real restriction, because experimentalists always rely on a vast amount of single runs and not only on one. The problem, however, is that Bob has to store his ensemble for the time Alice needs to produce $\rho_{Bob}$. Assume that the ensemble \{\{A$_i$\},\{B$_i$\}\} used to teleport the state of \{C$_i$\} consists of atoms or molecules. They can be kept, e. g., in an electromagnetic trap, but the field induces an interaction between the single entities so that Bob's \{B$_i$\} is more a multi-particle system than an ensemble in the strict sense, and it is to be expected that the interaction will influence the teleportation fidelity significantly. If, on the other hand, the B$_i$ are stored in a vessel, then this gas must be diluted so far that both dipolar and van der Waals interaction can be neglected and that during the storage time the collision probability is negligibly small.

It has, however, been shown \cite{a3} that the EPR correlations are \emph{contextual}. So, if the ensemble constituents interact regarding \emph{another} property type than the correlation is established, then storing in clouds or as a gas in a vessel becomes feasible. For example: If a molecule M$_\mathrm{2}$ with point group symmetry C$_\mathrm{i}$ dissociates into two chiral fragments M with opposite handedness, then it is to be expected that the fragments are EPR correlated (see \cite{a26} for a detailed description of this experiment), i. e., the correlation is established by chirality. This correlation is extremely stable against external fields and thermal collisions so that an interaction caused by the storage medium will not lead to decoherence (at least for the time necessary to store the ensemble). [This means that also the ensemble property is contextual. From a certain point of view a cloud of atoms is to be considered a multi-particle system whereas from another point of view it can behave as an ensemble in the strict sense.]
\item How to perform the required operation leading to the automatic impression of the unknown state $\vec{c}$ onto Bob's sub-ensemble? As long as an operation is unitary it is always possible to realize it experimentally \cite{a27}. But $\hat{P}_{aut}$ is self-adjoint, and it is an open question whether it corresponds to any experimentally accessible property type at all. On the other hand, however, $\hat{P}_{aut}$ represents a projection followed by a scaling-up procedure. Is this not a kind of re-normalization as assumed implicitly in \cite{a28}? Or must it be considered an actual ensemble multiplication which, in the case of photons, could be realized experimentally by a special photo-multiplier preserving the features of the incoming photons (see, e. g., \cite{a29})? [Note that a related problem arises in "general" quantum teleportation of a single particle where it could be resolved (partially) by adding an auxiliary particle on Bob's side only \cite{a30}.] In any case this points needs further investigation.
\end{itemize}
\section{Summary}
The feasibility of teleportation was examined from a pure ensemble point of view, and it has been shown that it is possible to teleport the state of an ensemble which is given by its statistical operator. While the sender has to apply physical realizations of projection operators the necessary operations on the side of the receiver are in general unitary.

It has been shown that teleportation allows for total abdication of any \textsc{a posteriori} information exchange between sender and receiver if they fix prior to the start of the experiment what Alice will do and when it shall happen.

For arbitrary incoming states complete ensemble teleportation is possible even if the receiver does not do anything with his sub-ensemble at all. In order to achieve said automatic teleportation the sender has to apply the physical realization of a special self-adjoint operator. The fidelity of this process amounts to 1.
\section*{Appendix}
\setcounter{equation}{0}
\renewcommand{\theequation}{\Roman{equation}}
Ansatz (16) is obviously justified, because the expectation value $\langle \hat{A} \rangle$ of \emph{any} property type A is given by
\begin{equation}
\langle \hat{A} \rangle = \mathrm{Tr}(\hat{A} \rho).
\end{equation}
One could, however, object that the preparation-induced change of the ensemble's statistical operator has to be reflected by the transformation
\begin{equation}
\rho_{total} \rightarrow \frac{\mathrm{Tr}_{\mathrm{C,A}} \left( (\hat{P} \otimes \hat{1}_B) \rho_{total} (\hat{P} \otimes \hat{1}_B) \right)}{\mathrm{Tr}_{\mathrm{total}}\left( (\hat{P} \otimes \hat{1}_B) \rho_{total} (\hat{P} \otimes \hat{1}_B) \right)}
\end{equation}
instead. For the numerator Tr$_{\mathrm{C,A}}(N)$ of the right hand side of (II) we obtain
\begin{eqnarray}
N &=& \frac{1}{8} \left( (\hat{C}_{11} \otimes \hat{A}_{11} + \hat{C}_{12} \otimes \hat{A}_{12} + \hat{C}_{21} \otimes \hat{A}_{21} + \hat{C}_{22} \otimes \hat{A}_{22}) \right. \nonumber \\ & & \left. \otimes \, (c_{11} \hat{B}_{22} - c_{12} \hat{B}_{21} - c_{21} \hat{B}_{12} + c_{22} \hat{B}_{11}) \right).
\end{eqnarray}
\begin{equation}
\Rightarrow \mathrm{Tr}_{\mathrm{C,A}}(N) = \frac{1}{4} \, (c_{11} \hat{B}_{22} - c_{12} \hat{B}_{21} - c_{21} \hat{B}_{12} + c_{22} \hat{B}_{11})
\end{equation}
Formation of the total trace yields the value $1/4 \, (c_{11} + c_{22}) = 1/4$ so that $\rho_{total}$ is finally mapped onto
\begin{equation}
c_{11} \hat{B}_{22} - c_{12} \hat{B}_{21} - c_{21} \hat{B}_{12} + c_{22}
\hat{B}_{11} = \sigma_3 \sigma_1 \rho_B \sigma_1 \sigma_3
\end{equation}
which is exactly the result (20) obtained in the usual way.

\end{document}